%% file: glob_journal.tex
\documentclass [journal,11pt]{IEEETran}

\makeatletter
\def\ps@headings{%
\def\@oddhead{\mbox{}\scriptsize\rightmark \hfil \thepage}%
\def\@evenhead{\scriptsize\thepage \hfil \leftmark\mbox{}}%
\def\@oddfoot{}%
\def\@evenfoot{}}
\makeatother
\usepackage[cmex10]{amsmath}
\usepackage[dvips]{graphicx}
\usepackage{subfigure}
\usepackage[small]{caption}
\usepackage{amsxtra}
\usepackage{algorithm}
\usepackage{soul}
\usepackage{ulem}
\usepackage{subfig}
\usepackage{algorithm, algorithmic}
\usepackage{latexsym,url,graphicx,amsmath,amssymb}
\usepackage{epsfig}
\usepackage{tabls}
\usepackage[dvips]{color}
\usepackage{times}
\usepackage{tabularx}
\usepackage{xspace}
\usepackage{pstricks}
\usepackage{multirow}
\usepackage{pst-grad}
\usepackage{theorem}

\newtheorem{theorem}{Theorem}[section]
\newtheorem{lemma}{Lemma}[section]

\newcommand{\BE}{\begin{equation}}
\newcommand{\EE}{\end{equation}}

\begin{document}

\title{\LARGE{Multiflow Transmission in Delay Constrained Cooperative Wireless Networks} }
\author{\IEEEauthorblockN{Marjan Baghaie, Dorit S. Hochbaum, Bhaskar Krishnamachari}\\
\thanks{Marjan Baghaie and Dorit S. Hochbaum are with University of California Berkeley and Bhaskar Krisnamachari is with University of Southern California. The work was conducted when all authors were with the University of Southern California.

*A preliminary version of the results presented in this paper appeared in IEEE Globecom conference \cite{Bagh11}.

This research was sponsored in part by the U.S. Army Research
Laboratory under the Network Science Collaborative Technology
Alliance, Agreement Number W911NF-09-2-0053, by NSF award No. DMI-0620677 and
CBET-0736232, and by NSF awards CNS-0627028 and CNS-1049541. Dorit S. Hochbaum is supported in part by NSF awards No. DMI-0620677, CMMI-1200592 and CBET-0736232.}
}
\maketitle

\input{abstract}
\input{multi}

\end{document}

%% file: abstract.tex
\begin{abstract}
This paper considers the problem of energy-efficient transmission in multi-flow multihop cooperative wireless networks. Although the performance gains of cooperative approaches are well known, the combinatorial nature of these schemes makes it difficult to design efficient polynomial-time algorithms for joint routing, scheduling and power control. This becomes more so when there is more than one flow in the network. It has been conjectured by many authors, in the literature, that the multiflow problem in cooperative networks is an NP-hard problem. In this paper, we formulate the problem, as a combinatorial optimization problem, for a general setting of $k$-flows, and formally prove that the problem is not only NP-hard but it is $o(n^{1/7-\epsilon})$ inapproxmiable. To our knowledge*, these results provide the first such inapproxmiablity proof in the context of multiflow cooperative wireless networks. We further prove that for a special case of k = 1 the solution is a simple path, and devise a polynomial
time algorithm for jointly optimizing routing, scheduling and power control. We then use this algorithm to establish analytical upper and lower bounds for the optimal performance for the general case of $k$ flows. Furthermore, we propose a polynomial time heuristic for calculating the solution for the general case and evaluate the performance of this heuristic under different channel conditions and against the analytical upper and lower bounds.
\end{abstract}
\IEEEpeerreviewmaketitle

%% file: multi.tex
\section{Introduction}

In a wireless network, a transmit signal intended for one node is received not only by that node but also by other nodes. In a traditional point-to-point system, where there is only one intended recipient, this innate property of the wireless propagation channel can be a drawback, as the signal constitutes undesired interference in all nodes but the intended recipient. However, this effect also implies that a packet \textit{can} be transmitted to multiple nodes simultaneously without additional energy expenditure. Exploiting this Òbroadcast advantageÓ, broadcast, multicast and multihop unicast systems can be designed to work cooperatively and thereby achieve potential performance gains. As such, cooperative transmission in wireless networks has attracted a lot of interest not only from the research community in recent years~\cite{Khandani03, Maric04, Mergen06, Mergen07, Jakllari07, Baghaie09, Baghaie11} but also from industry in the form of first practical cooperative mobile ad-hoc network systems~\cite{Halford10}. The majority of the work in the cooperative literature has so far focused on the single flow problem, though recently there has been an increased interest in considering multiflow settings in cooperative networks~\cite{infocom08, Ghaderi09, Ghaderi10, Azhang10,Azhang102, Bagh11}.

We consider a time-slotted system\footnote{Without loss of generality, we assume unit time-slots. Thus the terms energy and power are used interchangeably throughout this work.} in which the nodes that have received and decoded the packet are allowed to re-transmit it in future slots. During reception, nodes add up the signal power (EA)  received from multiple sources. Details of EA, and possible implementations have been extensively discussed in prior work~\cite{Maric04, Mergen06, Mergen07, Ghaderi09}. A key problem in such cooperative networks is routing and resource allocation, i.e., the question which nodes should participate in the transmission of data, and when, and with how much power, they should be transmitting.  The problem is further complicated when there is more than one flow going through the network at the same time.

We focus on the problem of minimum-energy multiflow cooperative transmission in this paper, where there are $k$ source-destination pairs, with each source node wanting to send a packet to its respective destination nodes, in a multihop wireless network. Other nodes in the network, that are neither a source nor a destination, may act as relays to help pass on the message through multiple hops, provided they have already decoded the message themselves and they are not transmitting/receiving any other messages at the time. The transmission is completed when all the destination nodes have successfully received their corresponding messages. It has been noted in the literature (\cite{Baghaie11,Baghaie12}) that a key tradeoff in cooperative settings is between the total energy consumption and the total delay measured in terms of the number of slots needed for all destination nodes in the network to receive the message. Therefore, we take delay into consideration and focus on the case where there is a delay constraint, whereby the destination node(s) should receive the message within some pre-specified delay constraint. We therefore formulate the problem of performing this transmission in such a way that the total transmission energy over all transmitting nodes is minimized, while meeting a desired delay constraint on the maximum number of slots that may be used to complete the transmission. The design variables in this problem determine which nodes should transmit, when, and with what power.

We furthermore assume that the nodes are memoryless, i.e., accumulation at the receiver is restricted to transmissions from multiple nodes in the present time slot, while signals from previous time slots are discarded. This assumption is justified (\cite{Baghaie11, Baghaie12}) by the limited storage capability of nodes in ad-hoc networks, as well as the additional energy consumption nodes have to expand in order to stay in an active reception mode when they ÒoverhearÓ weak signals in preceding time-slots.

The main contribution of the work presented in this chapter is as follows:  It has been \textit{conjectured} in the literature that \textit{the problem of jointly computing schedules, routing, and power allocation for multiple flows in cooperative networks} is NP-hard ~\cite{Azhang10,Azhang102, Ghaderi10}. In this chapter we formulate the joint problem of scheduling, routing and power allocation in a multiflow cooperative network setting and formally prove that not only it is NP-hard, but it is also $o(n^{1/7-\epsilon})$  inapproximable. (i.e., unless $P=NP$, it is not possible to develop a polynomial time algorithm for this problem that can obtain a solution that is strictly better than a logarithmic-factor of the optimum in all cases). We are not aware of prior work on multiflow cooperative networks that shows such inapproximability results. We further prove that for a special case of $k=1$, the solution is a simple path and devise an optimal polynomial time algorithm for joint routing, scheduling and power control. We establish analytical upper and lower bounds based on this algorithm and propose a polynomial-time heuristic, the performance of which is evaluated against those bounds.

The rest of this paper is organized as follows: In section \ref{sec:multi_formul} we provide a mathematical formulation of the problem. In section \ref{sec:multi_uni} we consider the special case of $k=1$ and prove the solution is a simple path and can be found optimally in polynomial time. The inapproximablity results are presented in section  \ref{sec:multi_hardness} using reduction from minimum graph coloring problem. We establish analytical upper and lower bounds for optimal performance in section \ref{sec:multi_bounds}. A polynomial-time heuristic is proposed in section \ref{sec:multi_heu} and its performance is evaluated under different channel conditions and against the performance bounds. Concluding remarks are summarized in section \ref{sec:multi_conc}.


\section{Problem Formulation} \label{sec:multi_formul}

 Consider a network, $G$, with a total of $n$ nodes, $I=\{1,..,n\}$. Assume we have $r$ source nodes, labeled ${\cal{S}} =\{s_1,s_2,...,s_r\}$, and $r$ corresponding destination nodes, ${\cal{D}} =\{d_1,d_2,...,d_r\}$. The source-destination nodes can be thought of as pairs, $\{(s_k,d_k)\}^r_{k=1}$, all with the same delay constraint $T$. The goal is to deliver a unicast message from each source to its corresponding destination, possibly using other nodes in the network as relays. The objective is to do so using the minimum amount of sum transmit power and within the delay constraint.

We consider a cooperative wireless setting with EA and consider signal-to-intereference-plus-noise (SINR) threshold model,  \cite{Ghaderi09,Maric04,BKTech, Ritesh}.  That is, in order for node $i$ to be able to decode message $k$ at time $t$, the following inequality needs to be satisfied:

\BE{\frac{\sum\limits_{j \in s_k(t)}{p_{jt}h_{ji}}}{\sum\limits_{u\notin s_k(t)}{p_{ut}h_{ui}}+N} \geq \theta .} \label{eq:condition}\EE
Here $s_k(t)$ is the set of nodes transmitting the message $k$ at time $t$, $h_{ij}$ is a constant between $0$ and $1$ representing the channel gain between node $i$ and $j$, and $N$ and $\theta$ are constants representing the noise and the decoding threshold respectively. 

Equation (\ref{eq:condition}) can be re-written as

\BE{\sum^{n}_{j =1}{h_{ji}p^k_{jt}}- \theta \sum^r_{\substack{ q=1\\
q \neq k}} \sum^{n}_{u = 1}{h_{ui}p^q_{ut}}- \theta N \geq 0,}\EE
where $p^k_{it}$ is the power used by node $i$ at time $t$ to transmit message $k$.

The system is memoryless, meaning although we are allowed to accumulate the same message from multiple sources during each time slot, we cannot accumulate over time. The relays are half-duplex, meaning they cannot transmit and receive simultaneously. The relays cannot transmit more than one message at the same time either.

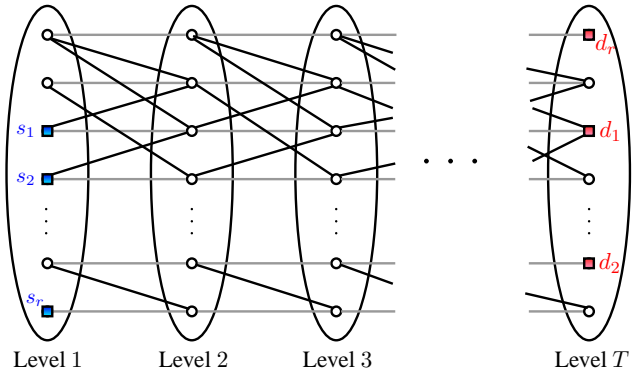
\begin{figure}[h!]
\center
\scalebox{0.8} 
{
\begin{pspicture}(0,-3.0425)(10.4,3.0025)
\definecolor{color308}{rgb}{0.6,0.6,0.6}
\definecolor{color1408g}{rgb}{1.0,0.2,0.2}
\definecolor{color1408f}{rgb}{1.0,0.6,0.8}
\pscircle[linewidth=0.04,dimen=outer](0.7,2.5025){0.1}
\pscircle[linewidth=0.04,dimen=outer](0.7,1.7025){0.1}
\pscircle[linewidth=0.04,dimen=outer](0.7,0.9025){0.1}
\pscircle[linewidth=0.04,dimen=outer](0.7,0.1025){0.1}
\psdots[dotsize=0.04](0.68,-0.3975)
\psdots[dotsize=0.04](0.68,-0.5975)
\psdots[dotsize=0.04](0.68,-0.7975)
\pscircle[linewidth=0.04,dimen=outer](0.7,-1.2975){0.1}
\pscircle[linewidth=0.04,dimen=outer](0.7,-2.0975){0.1}
\psellipse[linewidth=0.04,dimen=outer](0.7,0.2025)(0.7,2.8)
\pscircle[linewidth=0.04,dimen=outer](3.1,2.5025){0.1}
\pscircle[linewidth=0.04,dimen=outer](3.1,1.7025){0.1}
\pscircle[linewidth=0.04,dimen=outer](3.1,0.9025){0.1}
\pscircle[linewidth=0.04,dimen=outer](3.1,0.1025){0.1}
\psdots[dotsize=0.04](3.1,-0.3975)
\psdots[dotsize=0.04](3.1,-0.5975)
\psdots[dotsize=0.04](3.1,-0.7975)
\pscircle[linewidth=0.04,dimen=outer](3.1,-1.2975){0.1}
\pscircle[linewidth=0.04,dimen=outer](3.1,-2.0975){0.1}
\psellipse[linewidth=0.04,dimen=outer](3.1,0.2025)(0.7,2.8)
\pscircle[linewidth=0.04,dimen=outer](5.5,2.5025){0.1}
\pscircle[linewidth=0.04,dimen=outer](5.5,1.7025){0.1}
\pscircle[linewidth=0.04,dimen=outer](5.5,0.9025){0.1}
\pscircle[linewidth=0.04,dimen=outer](5.5,0.1025){0.1}
\psdots[dotsize=0.04](5.5,-0.3975)
\psdots[dotsize=0.04](5.5,-0.5975)
\psdots[dotsize=0.04](5.5,-0.7975)
\pscircle[linewidth=0.04,dimen=outer](5.5,-1.2975){0.1}
\pscircle[linewidth=0.04,dimen=outer](5.5,-2.0975){0.1}
\psellipse[linewidth=0.04,dimen=outer](5.5,0.2025)(0.7,2.8)
\pscircle[linewidth=0.04,dimen=outer](9.7,2.5025){0.1}
\pscircle[linewidth=0.04,dimen=outer](9.7,1.7025){0.1}
\pscircle[linewidth=0.04,dimen=outer](9.7,0.9025){0.1}
\pscircle[linewidth=0.04,dimen=outer](9.7,0.1025){0.1}
\psdots[dotsize=0.04](9.7,-0.3975)
\psdots[dotsize=0.04](9.7,-0.5975)
\psdots[dotsize=0.04](9.7,-0.7975)
\pscircle[linewidth=0.04,dimen=outer](9.7,-1.2975){0.1}
\pscircle[linewidth=0.04,dimen=outer](9.7,-2.0975){0.1}
\psellipse[linewidth=0.04,dimen=outer](9.7,0.2025)(0.7,2.8)
\psline[linewidth=0.04cm,linecolor=color308](0.8,2.5025)(3.0,2.5025)
\psline[linewidth=0.04cm,linecolor=color308](0.8,1.7025)(3.0,1.7025)
\psline[linewidth=0.04cm,linecolor=color308](0.8,0.9025)(3.0,0.9025)
\psline[linewidth=0.04cm,linecolor=color308](0.8,0.1025)(3.0,0.1025)
\psline[linewidth=0.04cm,linecolor=color308](0.8,-1.2975)(3.0,-1.2975)
\psline[linewidth=0.04cm,linecolor=color308](0.8,-2.0975)(3.0,-2.0975)
\psline[linewidth=0.04cm,linecolor=color308](3.2,2.5025)(5.4,2.5025)
\psline[linewidth=0.04cm,linecolor=color308](3.2,1.7025)(5.4,1.7025)
\psline[linewidth=0.04cm,linecolor=color308](3.2,0.9025)(5.4,0.9025)
\psline[linewidth=0.04cm,linecolor=color308](3.2,0.1025)(5.4,0.1025)
\psline[linewidth=0.04cm,linecolor=color308](3.2,-1.2975)(5.4,-1.2975)
\psline[linewidth=0.04cm,linecolor=color308](3.2,-2.0975)(5.4,-2.0975)
\psline[linewidth=0.04cm,linecolor=color308](5.6,2.5025)(6.5,2.5025)
\psline[linewidth=0.04cm,linecolor=color308](5.58,1.7025)(6.48,1.7025)
\psline[linewidth=0.04cm,linecolor=color308](5.58,0.9025)(6.48,0.9025)
\psline[linewidth=0.04cm,linecolor=color308](5.58,0.1025)(6.48,0.1025)
\psline[linewidth=0.04cm,linecolor=color308](5.58,-1.2975)(6.48,-1.2975)
\psline[linewidth=0.04cm,linecolor=color308](5.58,-2.0975)(6.48,-2.0975)
\psline[linewidth=0.04cm,linecolor=color308](9.6,2.5025)(8.7,2.5025)
\psline[linewidth=0.04cm,linecolor=color308](9.6,1.7025)(8.7,1.7025)
\psline[linewidth=0.04cm,linecolor=color308](9.6,0.9025)(8.7,0.9025)
\psline[linewidth=0.04cm,linecolor=color308](9.6,0.1025)(8.7,0.1025)
\psline[linewidth=0.04cm,linecolor=color308](9.6,-2.0975)(8.7,-2.0975)
\psline[linewidth=0.04cm,linecolor=color308](9.6,-1.2975)(8.7,-1.2975)
\psline[linewidth=0.04cm](0.76,2.4425)(3.04,1.7625)
\psline[linewidth=0.04cm](0.74,2.4425)(3.06,0.9625)
\psline[linewidth=0.04cm](0.72,0.9625)(3.04,1.6425)
\psline[linewidth=0.04cm](0.72,1.6425)(3.04,0.1625)
\psline[linewidth=0.04cm](0.72,0.1625)(3.04,0.8825)
\psline[linewidth=0.04cm](0.74,-1.3375)(3.06,-2.0575)
\psline[linewidth=0.04cm](3.12,2.4825)(5.44,1.7625)
\psline[linewidth=0.04cm](3.16,0.9625)(5.44,1.6425)
\psline[linewidth=0.04cm](3.12,2.4425)(5.44,0.9625)
\psline[linewidth=0.04cm](3.16,1.6825)(5.44,0.1625)
\psline[linewidth=0.04cm](3.16,0.1625)(5.44,0.8425)
\psline[linewidth=0.04cm](3.16,-1.3175)(5.44,-2.0375)
\psline[linewidth=0.04cm](5.56,2.4825)(6.44,2.2025)
\psline[linewidth=0.04cm](5.56,2.4425)(6.48,1.9225)
\psline[linewidth=0.04cm](5.52,1.6425)(6.44,1.2825)
\psline[linewidth=0.04cm](5.56,0.9625)(6.44,1.1225)
\psline[linewidth=0.04cm](5.56,0.1625)(6.48,0.3625)
\psline[linewidth=0.04cm](5.56,-1.3175)(6.44,-1.6375)
\psline[linewidth=0.04cm](9.62,1.7425)(8.66,1.9425)
\psline[linewidth=0.04cm](9.64,1.6825)(8.72,1.4025)
\psline[linewidth=0.04cm](9.64,0.9625)(8.76,1.2825)
\psline[linewidth=0.04cm](9.64,0.8425)(8.76,0.4025)
\psline[linewidth=0.04cm](9.64,0.1625)(8.68,0.6425)
\psline[linewidth=0.04cm](9.68,-2.0375)(8.64,-1.7975)
\psdots[dotsize=0.08](7.0,0.4025)
\psdots[dotsize=0.08](7.4,0.4025)
\psdots[dotsize=0.08](7.8,0.4025)
\psframe[linewidth=0.04,dimen=outer,fillstyle=gradient,gradlines=2000,gradmidpoint=1.0](0.8,1.0025)(0.6,0.8025)
\psframe[linewidth=0.04,dimen=outer,fillstyle=gradient,gradlines=2000,gradmidpoint=1.0](0.8,0.2025)(0.6,0.0025)
\psframe[linewidth=0.04,dimen=outer,fillstyle=gradient,gradlines=2000,gradmidpoint=1.0](0.8,-1.9975)(0.6,-2.1975)
\psframe[linewidth=0.04,dimen=outer,fillstyle=gradient,gradlines=2000,gradbegin=color1408g,gradend=color1408f,gradmidpoint=1.0,fillcolor=color1408g](9.8,2.6025)(9.6,2.4025)
\psframe[linewidth=0.04,dimen=outer,fillstyle=gradient,gradlines=2000,gradbegin=color1408g,gradend=color1408f,gradmidpoint=1.0,fillcolor=color1408g](9.8,1.0025)(9.6,0.8025)
\psframe[linewidth=0.04,dimen=outer,fillstyle=gradient,gradlines=2000,gradbegin=color1408g,gradend=color1408f,gradmidpoint=1.0,fillcolor=color1408g](9.8,-1.1975)(9.6,-1.3975)
\usefont{T1}{ptm}{m}{n}
\rput(0.714375,-2.8875){Level $1$}
\usefont{T1}{ptm}{m}{n}
\rput(3.1289062,-2.8875){Level $2$}
\usefont{T1}{ptm}{m}{n}
\rput(5.521406,-2.8875){Level $3$}
\usefont{T1}{ptm}{m}{n}
\rput(9.751875,-2.8875){Level $T$}
\usefont{T1}{ptm}{m}{n}
\rput(0.3459375,0.9025){\small \color{blue}$s_1$}
\usefont{T1}{ptm}{m}{n}
\rput(0.3456875,0.1025){\small \color{blue}$s_2$}
\usefont{T1}{ptm}{m}{n}
\rput(0.50265625,-1.8975){\small \color{blue}$s_r$}
\usefont{T1}{ptm}{m}{n}
\rput(10.05625,0.9125){\color{red}$d_1$}
\usefont{T1}{ptm}{m}{n}
\rput(10.050781,-1.2875){\color{red}$d_2$}
\usefont{T1}{ptm}{m}{n}
\rput(9.983282,2.3525){\color{red}$d_r$}
\end{pspicture}}
\caption{Applying the multicommodity flow technique for unicast cast} 
\label{fig:GF}
\end{figure}

In order to apply ideas driven by the rich literature on multicommodity flows~\cite{Hoch96} to our problem, we need to somehow introduce the notion of delay constraint into the multicommodity setting. What follows is a transformation of our network graph that would allow for the multicommodity flow technique to be applied, while observing the delay constraint: For a delay constraint $T$, map the given network to a layered graph with $T$ layers as shown in Figure \ref{fig:GF}. Place a copy of all the nodes in the network on each of the layers. Connect each node, on each layer, to its corresponding copy on its neighboring layers with an edge weight of $0$. Also create directed edges between each node, on each layer $k$, and the nodes on the next layer $k+1$, with edge weights representing the amount of power required to transmit the message from the node on the top level to the node on the bottom level, as a whole. Notice that there is no edge between the nodes on the same level.  Call the new graph $G'$.  Assign the nodes corresponding to the source nodes of $G$ on level $1$ of $G'$ as source nodes in $G'$ and the destination nodes on level $T$ of $G'$, corresponding to destination nodes in $G$, as destinations in $G'$, as shown in the figure. Similar transformations have been used in the literature in the context of multiflow transmission \cite{Azhang10}.

Without loss of generality, we assume unit length time slots. The nodes who want to transmit are to do so at the beginning of each time slot, and the decoding (by nodes who receive enough information during that time slot) will happen by the end of that time slot. Let $z^k_{it}$ be an indicator binary variable that indicates whether or not node $i$ decodes the message $k$ during time slot $t$, as per inequality in equation (\ref{eq:condition}). In other words, we define $z^k_{it}$ to be $1$, if node $i$ decodes message $k$ during time slot $t$, and $0$ otherwise.  Let $p^k_{it}$ be the transmit power used by node $i$ at each time $t$ to transmit message $k$. We define another binary variable $x^k_{it}$, that is $1$ if node $i$ is {\it{allowed}} to transmit message $k$ at time $t$, and $0$ otherwise. A node is {\it{allowed}} to transmit during a particular time slot, if it has already decoded that message in previous time slots, and it's not receiving or transmitting any other messages during that time slot. Notice that being allowed to transmit does not necessarily mean that a transmission actually occurs. To take care of actual transmissions, let us define $v^k_{it}$ to be a binary variable that is $1$ if node $i$ transmits message $k$ at time $t$, and $0$ otherwise.

The problem can then be formalized as a combinatorial optimization problem:

\begin{eqnarray} \label{eqn:MF} \min & P_{total} = \sum_{t=1}^{T}\sum_{i=1}^{n}\sum_{k=1}^{r}  p^k_{it} \\  \begin{tabular}{r}
  s.t. \\
 \\
    \\
\\
   \\
\\
    \\
\\
   \\
\\
\end{tabular}
& \begin{tabular}{ >{$}l<{$} l }
  1. &$ p^k_{it} \geq 0,~~\forall i , t, k$   \\
  2. &$ x^k_{d_kT+1} =1,~~\forall k$   \\
  3. & $x^k_{it+1} \leq  z^k_{it}+ x^k_{it},~~\forall i, t$  \\
  4. & $(-M)(1-z^k_{it}) \leq y^k_{it},~~\forall i,  t$  \\
  5. & $p^k_{it} \leq Mv^k_{it},~~\forall i, t$  \\
  6. & $\sum_{k=1}^{r}\left({v^k_{it}+z^k_{it}}\right) \leq 1,~~\forall i, t$  \\
  7. &$ v^k_{it} \leq x^k_{it},~~\forall i , t, k$   \\
 8. & $x^k_{s_k1} =z^k_{s_k1} =1, \forall k$ \\
   9. & $x^k_{i1} =z^k_{i1} =0, \forall i \in I\backslash\{s_k\}$\\
10. & $x^k_{it} \in \{0,1\}  $\\
11. & $z^k_{it} \in \{0,1\}  $\\
12. & $v^k_{it} \in \{0,1\} . $\\

\end{tabular} \nonumber \end{eqnarray}
Here $y^k_{it} =\sum^{n}_{j =1}{h_{ji}p^k_{jt}}- \theta \sum^r_{\substack{ q=1\\
q \neq k}} \sum^{n}_{u = 1}{h_{ui}p^q_{ut}}- \theta N$, $M$ is a large positive constant, and the constraints have the following interpretations:

\begin{enumerate}

\item No negative power is allowed.
\item Every node in the destination set is required to have decoded the data by the end of time slot $T$.
\item If a node has not decoded a message by the end of time slot $t$, that node is not allowed to transmit that message at time $t+1$.
\item $z^k_{ti}$ is forced to be  $0$ if message $k$ is not decoded in time slot $t$.
\item $p^k_{it}$ is forced to be  $0$, if node $i$ is not transmitting message $k$ at time $t$ (i.e. if $v^k_{it} = 0$).
\item A node cannot transmit and receive at the same time and can only transmit or receive a single message at each time slot.
\item $v^k_{it}$ is forced to be $0$, node $i$ is not allowed to transmit message $k$ at time $t$ (i.e. if $x^k_{it} = 0$).
\item Only sources have the message at the beginning.
\item No one else has the message at the beginning.
\item $x$, $z$ and $v$ are binary variables.

\end{enumerate}
We call this optimization problem MCUE, for multiflow cooperative unicast with Energy Accumulation.

\section{Special Case of $k=1$} \label{sec:multi_uni}
In this section we consider MCUE for the special case of $k=1$ and prove the problem can be solved optimally and in polynomial time for this special case. We also provide a polynomial-time algorithm to achieve the optimum solution.

\begin{theorem}{ The optimal solution for MCUE is a simple path for $k=1$, but not necessarily so for $k>2$.}
\end{theorem} \label{them:multi_uni}
\begin{proof} {The claim can be proved by induction on $T$: For delay $T=1$, the claim is trivially true, as the optimal solution is direct transmission from the source, $s$, to the given destination, $d$. Let us assume the claim is true for $T=t-1$. To complete the proof, we need to show the claim holds for $T=t$. Pick any node in the network as the desired destination $d$. If the message can be transmitted from source $s$ to $d$ with minimum energy in a time frame less than $t$, then an optimal simple path exists by the induction assumption. So consider the case when it takes exactly $T=t$ steps to turn on $d$. The system is memoryless, so $d$ must decode by accumulating the energy transmitted from a set of nodes, $\textbf{v}$, at time $t$. This can be represented as $\sum\limits_{v_i\in \textbf{v}} p_{{v_i}t}h_{dv_i} \geq \theta$. We observe that there must exist a node $v_o \in \textbf{v}$ whose channel to $d$ is equal or better than all the other nodes in $\textbf{v}$. Therefore, given $h_{dv_o} \geq h_{dv_i}, \forall v_i \in \textbf{v}\backslash\{v_o\}$ then $\sum\limits_{v_i\in \textbf{v}} p_{{v_i}t}h_{dv_o} \geq \sum\limits_{v_i\in \textbf{v}} p_{{v_i}t}h_{dv_i} \geq \theta$. In other words, if we add the power from all nodes in $\textbf{v}$ and transmit instead from $v_o$, our solution cannot be worse. $v_o$ must have received the message by time $t-1$, to be able to transmit the message to $d$ at time $t$. We know by the induction assumption that the optimal simple path solution exists from source to any node to deliver the message within $t-1$ time frame. Thus, for $T=t$, there exists a simple path solution between $s$ and $d$, which is optimum.}\end{proof} Considering the above theorem, the MCUE problem formulation (for the special case of $k=1$) reduces to:
\begin{eqnarray} \label{eqn:MFk1} \min& \hspace{-4cm}  P_{total} = \sum_{t=1}^{T}\sum_{i=1}^{n}  p_{it}  \\
\begin{tabular}{r}
  s.t.
     \\
\\
   \\
\\
    \\
\\
   \\
\\
\end{tabular}
& \begin{tabular}{ >{$}l<{$} l }
  1. &$ p_{it} \geq 0,~~\forall i , t$   \\
  2. &$ x_{dT+1} =1$   \\
  3. &\small{$-M(1-x_{it+1}) \leq \sum\limits^{n}_{j =1}{h_{ji}p_{jt}}-\theta N,~~\forall i,  t$ } \\
  4. & $p_{it} \leq Mx_{it},~~\forall i, t$  \\
 5. & $x_{s1}  =1$ \\
   6. & $x_{i1}  =0, \forall i \neq s$\\
7. & $x_{it} \in \{0,1\}  $\\
\end{tabular} \nonumber \end{eqnarray}

This can be solved optimally in polynomial time using dynamic programming. Let  $C(i,t)$ be the minimum cost it takes for source node $s$ to turn on $i$, possibly using relays, within at most $t$ time slots. Then we can write:
\BE{ C(i,t) =\min_{j \in Nr(i)}\left[C(j,t-1)+w_{ji}\right] }\\\EE \label{eq.multi_uni} with $C(s,t) = 0$, for all $t$ and  $C(i,1) = w_{si}$, where  $Nr(i)$ is the set that contains $i$ and its neighboring nodes that have a non-zero channel to $i$, $w_{ji}$ represents the power it takes for $j$ to turn on $i$ using direct transmission. Thus the solution to (\ref{eqn:MFk1}) is given by $C(d,T)$ and its computation incurs a running time of $O(n^3)$.


\section{Inapproximability Results} \label{sec:multi_hardness}

\begin{figure}[htb!]
\center

\scalebox{1} 
{
\begin{pspicture}(0,-1.3)(8.88,1.3)
\definecolor{color308}{rgb}{0.6,0.6,0.6}
\definecolor{color1603b}{rgb}{1.0,0.2,0.2}
\definecolor{color1605b}{rgb}{1.0,0.0,0.2}
\pscircle[linewidth=0.04,dimen=outer](0.8240625,1.105){0.1}
\pscircle[linewidth=0.04,dimen=outer](0.8840625,-0.015){0.1}
\pscircle[linewidth=0.04,dimen=outer](0.8440625,-1.035){0.1}
\psline[linewidth=0.04cm](0.9240625,1.105)(3.1240625,1.105)
\psline[linewidth=0.04cm](0.9840625,-0.015)(3.1840625,-0.015)
\psline[linewidth=0.04cm](0.9440625,-1.035)(3.1440625,-1.035)
\usefont{T1}{ptm}{m}{n}
\rput(0.43975,-1.075){\small \color{blue}$s_3$}
\usefont{T1}{ptm}{m}{n}
\rput(0.5,-0.075){\small \color{blue}$s_2$}
\usefont{T1}{ptm}{m}{n}
\rput(0.42,1.065){\small \color{blue}$s_1$}
\pscircle[linewidth=0.04,dimen=outer](3.1840625,1.105){0.1}
\pscircle[linewidth=0.04,dimen=outer](3.2640624,-0.015){0.1}
\pscircle[linewidth=0.04,dimen=outer](3.2040625,-1.035){0.1}
\pscircle[linewidth=0.04,dimen=outer](5.5840626,1.105){0.1}
\pscircle[linewidth=0.04,dimen=outer,fillstyle=solid,fillcolor=color1603b](5.6640625,0.625){0.1}
\pscircle[linewidth=0.04,dimen=outer,fillstyle=solid,fillcolor=color1605b](5.6840625,-0.615){0.1}
\pscircle[linewidth=0.04,dimen=outer](5.6040626,-1.035){0.1}
\psline[linewidth=0.04cm](3.2840624,1.105)(5.4840627,1.105)
\psline[linewidth=0.04cm](3.3040626,-1.035)(5.5040627,-1.035)
\pscircle[linewidth=0.04,dimen=outer](7.9840627,1.105){0.1}
\pscircle[linewidth=0.04,dimen=outer](8.004063,0.025){0.1}
\pscircle[linewidth=0.04,dimen=outer](8.004063,-1.035){0.1}
\psline[linewidth=0.04cm](5.6840625,1.105)(7.8840623,1.105)
\psline[linewidth=0.04cm](5.7040625,-1.035)(7.9040623,-1.035)
\psline[linewidth=0.04cm,linecolor=color308](5.75975,-0.595)(7.91975,-0.975)
\psline[linewidth=0.04cm,linecolor=color308](5.73975,0.605)(7.87975,1.085)
\usefont{T1}{ptm}{m}{n}
\rput(8.4,1.105){\small \color{blue}$d_1$}
\usefont{T1}{ptm}{m}{n}
\rput(8.4,0.025){\small \color{blue}$d_2$}
\usefont{T1}{ptm}{m}{n}
\rput(8.4,-1.035){\small \color{blue}$d_3$}
\psdiamond[linewidth=0.04,dimen=outer](5.65975,-0.0050)(2.32,0.57)
\end{pspicture}
}
\caption{An example of $k>2$, with $T=3$, where the optimal solution is not a single path.} 
\label{fig:multik}
\vspace{0.5in}
\end{figure}
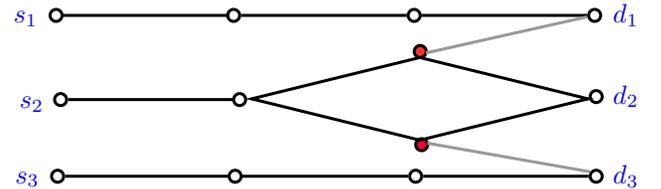

For $k=1$, we proved in Theorem \ref{them:multi_uni}, that the optimal solution is a simple path. For $k>2$, we can consider the following counter-example to argue that the solution is not necessarily a single-path. Consider the scenario shown in Figure \ref{fig:multik}, where $T=3$, where the edge weights are equal and the edges shown in gray show strong interference. The red nodes cannot by themselves transmit the message to $d_2$, as it causes interference for $d_1$ and $d_3$ preventing them from being able to decode the data. However, they can cooperate with each other, by each sending with half power to get the message to $d_2$ without causing too much interference for the other destinations.

To investigate the complexity of MCUE, let us start by looking at a sub-problem. Imagine a one hop setting of $k$ source nodes and their corresponding $k$ destination nodes, with no relay nodes. Due to interference, not all sources can transmit simultaneously. The task is to schedule the sources appropriately, so that everyone can get their message delivered to their corresponding destination within a time delay $T$. The problem is to find the minimum such $T$. Let us call this problem MOSP, for multi-source one-hop scheduling problem\footnote{This is essentially the problem considered in \cite{Tam}, though no proof of complexity is given in that paper.}. It is important to note that MCUE is at least as hard as MOSP. Thus, any hardness results obtained for MOSP imply hardness of MCUE.

In this section, we derive inapproximablity results for MOSP by showing that any instance of \textit{minimum graph coloring problem} \cite{Hoch96} can be reduced to an instance of MOSP.

\begin{lemma} \label{lem.mosp_hard} MOSP is $o(n^{1/7-\epsilon})$ inapproximable, for any $\epsilon > 0$. \end{lemma}
\begin{proof} Given an instance $G(V,E)$, $|V| =n$, of the minimum graph coloring, we construct a bipartite graph $G'$, with the bi-partition $X$ and $Y$ with  $|X|=|Y|=n$. For each node $v_i \in G$, we place two nodes $u_i \in X$ and $u'_i \in Y$ and connect them with an edge $(u_i,u'_i)$. Also for every edge in $G$, $e_{ij}=\{v_i,v_j\}$, place two edges $(u_i,u'_j)$ and $(u_j,u'_i)$ in $G'$.  We assign $u_i$ and $u'_i$ to be a source and destination pair respectively for all $i$. We set equal edge weights for all the edges in $G'$ and set $\theta >1$ to get an instance of MOSP.

A simple example is shown in Figure \ref{fig:Multicolor}. Notice that the gray edges in the figure represent interference, and by setting $\theta>1$, a message can be successfully decoded if and only if there is no interference at that node.

\begin{figure}[htb!]
\hspace{-0.5cm}
\scalebox{1} 
{
\begin{pspicture}(0,-1.84)(9.353281,1.84)
\definecolor{color308}{rgb}{0.6,0.6,0.6}
\pscircle[linewidth=0.04,dimen=outer](5.9973435,1.045){0.1}
\pscircle[linewidth=0.04,dimen=outer](5.9973435,0.245){0.1}
\pscircle[linewidth=0.04,dimen=outer](5.9973435,-0.555){0.1}
\pscircle[linewidth=0.04,dimen=outer](5.9973435,-1.355){0.1}
\pscircle[linewidth=0.04,dimen=outer](8.397344,1.045){0.1}
\pscircle[linewidth=0.04,dimen=outer](8.397344,0.245){0.1}
\pscircle[linewidth=0.04,dimen=outer](8.397344,-0.555){0.1}
\pscircle[linewidth=0.04,dimen=outer](8.397344,-1.355){0.1}
\psline[linewidth=0.04cm](6.097344,1.045)(8.297344,1.045)
\psline[linewidth=0.04cm](6.097344,0.245)(8.297344,0.245)
\psline[linewidth=0.04cm](6.097344,-0.555)(8.297344,-0.555)
\psline[linewidth=0.04cm](6.097344,-1.355)(8.297344,-1.355)
\psline[linewidth=0.04cm,linecolor=color308](6.057344,0.985)(8.337344,0.305)
\psline[linewidth=0.04cm,linecolor=color308](6.037344,0.985)(8.313031,-1.295)
\psline[linewidth=0.04cm,linecolor=color308](6.0730314,-0.555)(8.337344,0.185)
\psline[linewidth=0.04cm,linecolor=color308](6.0730314,0.245)(8.313031,-0.535)
\usefont{T1}{ptm}{m}{n}
\rput(5.6132812,-0.595){\small \color{blue}$u_3$}
\usefont{T1}{ptm}{m}{n}
\rput(5.6130314,-1.395){\small \color{blue}$u_4$}
\pscircle[linewidth=0.04,dimen=outer](2.0770316,1.025){0.1}
\pscircle[linewidth=0.04,dimen=outer](0.8170315,-0.155){0.1}
\pscircle[linewidth=0.04,dimen=outer](2.0770316,-1.335){0.1}
\pscircle[linewidth=0.04,dimen=outer](3.3570316,-0.155){0.1}
\usefont{T1}{ptm}{m}{n}
\rput(2.08,1.305){\small \color{blue}$v_1$}
\usefont{T1}{ptm}{m}{n}
\rput(0.44,-0.115){\small \color{blue}$v_2$}
\usefont{T1}{ptm}{m}{n}
\rput(2.08,-1.615){\small \color{blue}$v_3$}
\usefont{T1}{ptm}{m}{n}
\rput(3.72,-0.115){\small \color{blue}$v_4$}
\usefont{T1}{ptm}{m}{n}
\rput(5.633281,0.205){\small \color{blue}$u_2$}
\usefont{T1}{ptm}{m}{n}
\rput(5.6132812,1.005){\small \color{blue}$u_1$}
\psline[linewidth=0.04cm,linecolor=color308](6.0730314,0.245)(8.293032,1.025)
\psline[linewidth=0.04cm,linecolor=color308](6.0730314,-0.575)(8.293032,-1.335)
\psline[linewidth=0.04cm,linecolor=color308](6.0930314,-1.355)(8.293032,1.025)
\psline[linewidth=0.04cm,linecolor=color308](6.0730314,-1.355)(8.333032,-0.555)
\usefont{T1}{ptm}{m}{n}
\rput(8.843281,1.045){\small \color{blue}$u'_1$}
\usefont{T1}{ptm}{m}{n}
\rput(8.823281,0.245){\small \color{blue}$u'_2$}
\usefont{T1}{ptm}{m}{n}
\rput(8.823281,-0.555){\small \color{blue}$u'_3$}
\usefont{T1}{ptm}{m}{n}
\rput(8.823281,-1.355){\small \color{blue}$u'_4$}
\usefont{T1}{ptm}{m}{n}
\rput(2.05,1.645){\small $G$}
\usefont{T1}{ptm}{m}{n}
\rput(7.2,1.625){\small $G'$}
\psdiamond[linewidth=0.04,dimen=outer](2.0830312,-0.155)(1.21,1.12)
\end{pspicture}
}

\caption{Example construction of $G'$, for a given $G$.} 
\label{fig:Multicolor}
\end{figure}
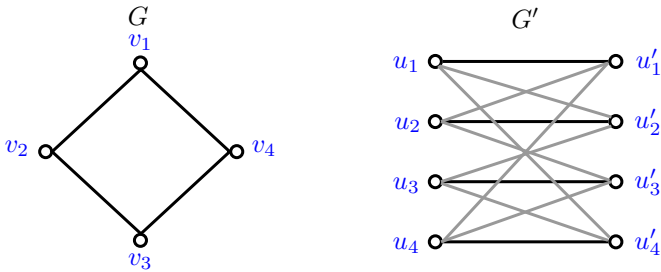

\end{proof}
This in turn means two sources in $G'$ can simultaneously transmit if and only if there is no edge in between them in $G$. Thus, the set of nodes that are transmitting simultaneously in $G'$ correspond to an independent set in $G$. Consequently, the optimal solution to MOSP is equal to the minimum graph coloring of $G$, which is known to be $o(n^{1/7-\epsilon})$ inapproximable \cite{Bell}.
The following theorem follows by noticing that MOSP is a special case of MCUE.
\begin{theorem} \label{thm.mcue_hard} MCUE is $o(n^{1/7-\epsilon})$ inapproximable,for any $\epsilon > 0$. \end{theorem}

Notice that the inapproximability result, given by Theorem \ref{thm.mcue_hard}, is stronger than, and implies, the NP-hardness result. In other words, it implies that not only finding the optimal solution is NP-hard but finding a polynomial time approximation algorithm that approximates the optimal solution to MCUE with a factor of $o(n^{1/7-\epsilon})$ is also NP-hard.


\section{Performance Bounds} \label{sec:multi_bounds}

In section  \ref{sec:multi_hardness}, we proved that MCUE problem is in general inapproximable. However, it was shown in section \ref{sec:multi_uni} that the problem can be solved optimally and in polynomial time for the special case of $k=1$. In this section, we use the results of section \ref{sec:multi_uni} to obtain performance bounds for MCUE.

\subsection{An Analytical Lower Bound}\label{sec:multi_lowerbounds}

In this section we establish a lower bound on the optimum solution to MCUE.

To get a better intuition for this lower bound, let us start off by considering the optimal solution to MCUE for the case when there is only one flow present in the network. As before, we have $n$ nodes and a channel $H$, but this time the source $s$ wants to transmit the message to a particular destination $d$, using the minimum energy within a given delay constraint $T$. The system is cooperative in that other nodes in the network, may be utilized as memoryless energy accumulating relays to help achieve the minimum energy goal.  Based on section  \ref{sec:multi_uni}, the solution can be found by calculating $C(d,T)$ where $C(d,T)$ is defined as per equation  (\ref{eq.multi_uni}).

To find a lower bound for MCUE for a general case of $r$ flows, with source-destination pairs $\{(s_k,d_k)\}^r_{k=1}$, all with the same delay constraint $T$, we notice that the cost paid by optimal MCUE to cover each node cannot be lower than the optimal minimum cost paid by each source $s_k$ to cover its corresponding destination $d_k$ in the absence of other interfering flows. Based on that observation we derive the following lower-bound, $LB(T)$, for the $OPT_{MCUE}$ for $r$ flows when the delay constraint is $T$:
\BE {LB(T) = \sum_{k=1}^{r}  C(d_k,T)} \EE
where $C(d_k,T)$ is defined as per equation  (\ref{eq.multi_uni}). In other words, $C(d_k,T)$ calculates the minimum cost of optimal single flow transmission to cover a destination $d_k$, starting from its corresponding source under a delay constraint $T$. $LB(T)$ takes the sum of those costs and use it as lower-bound - since we know $OPT_{MCUE}$ has to cover all these flows and cannot do so any better than the optimal solution for a single flow. Computing this lower bound incurs a running time of $O(n^3)$.

\subsection{An Analytical Upper Bound}\label{sec:multi_upperbound}

In this section we establish an upper bound on the optimum solution to MCUE, for the general case of $r$ flows, with $T \geq r$.

The upper bound is established by considering the multiplexing solution. At the extreme end of $T = r$, we would allow one time slot for each of the $r$ flow to transmit its message, while the other flows are silent.  For a general time $T (> r)$ we break the time into $r$ blocks $\mathcal{T} = (\tau_1, \tau_k,...\tau_r)$, such that $\sum_{k=1}^{r} \tau_k = T$. We assign each block to one of the flows, while the other flows are silent. We calculate $C(d_k,\tau_k)$, defined as per equation  (\ref{eq.multi_uni}). For a given tuple $\mathcal{T}$, the summation of the total energy required by all flows to complete their transmission can be achieved by calculating:
\BE UB(\mathcal{T})=\sum_{k=1}^{r}C(d_k,\tau_k) \EE This sum would provide an upper bound for $OPT_{MCUE}$.  For a general $T \leq r$, we will have ${T-1 \choose k-1}$ possibilities for assigning the time slots to different flows. The upper bound is calculated as follows:

\BE UB(T) = \min_\mathcal{T} UB(\mathcal{T})\EE \label{eq.multi_ub}

To compute this upper bound we need to carry on the computation for calculating a single flow MCUE, discussed in section \ref{sec:multi_uni}, $ {T-1 \choose k-1} \times r $ times. Thus the upper bound incurs a running time of $O(n^3)$.


\section{A Polynomial-time Heuristic} \label{sec:multi_heu}

In this section we propose a polynomial time heuristic for MCUE, the performance of which is later evaluated against that of the bounds established in section \ref{sec:multi_bounds}. We call this polynomial time heuristic MCUH, for multiflow cooperative unicast heuristic.

 To recap, consider a network, $G$, with a total of $n$ nodes, $I=\{1,..,n\}$. Assume we have $r$ source nodes, labeled ${\cal{S}} =\{s_1,s_2,...,s_r\}$, and $r$ corresponding destination nodes, ${\cal{D}} =\{d_1,d_2,...,d_r\}$. The source-destination nodes can be thought of as pairs, $\{(s_k,d_k)\}^r_{k=1}$, all with the same delay constraint $T$. The goal is to deliver a unicast message from each source to its corresponding destination, possibly using other nodes in the network as relays. The objective is to do so using the minimum amount of sum transmit power and within a given delay constraint. 
 
The MCUH algorithm works greedily by scheduling flows one by one.  Each flow is given more slots than its previous flows, to ensure a feasible solution always exist. That means the algorithm works for $T \geq r$. Each flow, with the exception of the final flow, uses more power than required to deliver its message. This is achieved by assigning a higher threshold to that flow when scheduling the flow. After scheduling, the nodes that will be transmitting at each time slot and the power they use for transmission is passed on to the next flow. Each flow, when scheduling itself, will ensure that its transmission will not disturb the transmission of previously scheduled flows. A lower threshold is used to check for disturbance, than the one used for scheduling the flow itself. Let us now look at the details of the algorithm.

We schedule the $r$ flows greedily, starting from the one that causes the least disturbance. Without loss of generality, let us assume that we are scheduling the flows in the order $1$ to $r$. All flows need to be scheduled within a total of $T$ time slots. For flow $1$, we assign a delay constraint of $T_1$  time slots for transmission, for flow $2$, we assign $T_2$ and so forth, such  that: \BE {1 \leq T_1<T_2<...<T_r= T}\EE Recall that time-slots are in unit durations, thus can increment in integer units. Therefore, each flow has at least one more time slot at its disposal than its immediate predecessor, ensuring that a feasible solution always exists. This also means that the algorithm works for $T \geq r$.

In section \ref{sec:multi_formul},  we defined $\theta$ to be the decoding threshold as per equation (\ref{eq:condition}). For this multi-flow setting, each flow is assigned its own $\theta$ value, such that: \BE{\theta_1 > \theta_2 > ...> \theta_r= \theta}\EE

Flow $1$ is scheduled with $T_1$ and $\theta_1$, as per algorithm in section \ref{sec:multi_uni}.  We store the nodes that are scheduled to transmit in each time slot, and their transmit power and their corresponding receivers in a black list $\mathcal{B}$. According to this definition, $\mathcal{B}(t)$ gives us the set of already scheduled nodes that are transmitting at time $t$ and their corresponding powers, and their corresponding receiving nodes.

For the $k$th flow, we use a modified version of the DMECT-go algorithm discussed in \cite{Baghaie11, Baghaie12}. DMECT-go is a polynomial time algorithm that uses a deterministic dynamic program to optimally solve the problem of joint scheduling and power allocation, for any given ordering, in a single-flow network. In \cite{Baghaie11, Baghaie12}, it was shown that Dijkstra's shortest path algorithm provides a good heuristic for ordering in a network with uniform distribution of nodes.

As mentioned, DMECT-go is for a single source problem and does not take interference or deliverance of multiple messages from multiple sources into account.  The algorithm proposed in this section, MCUH, is a modified version of DMECT-go. In MCUH, we broke the NP hard problem of MCUE into three subproblems namely, ordering, scheduling and power allocation.  Ordering, for a vector of n nodes, is defined as an array of indices from $1$ to $n$; any node that has decoded the message will only be allowed to retransmit when all nodes with smaller index have also decoded the message (and are thus allowed to take part in transmission). Given ordering, what remains to be determined is scheduling, and power allocation. In other words, what remains is deciding which nodes should take part in transmission of each flow, how much power they should transmit with and at what time slots, such that minimum energy is consumed while delay constraints are satisfied.  We also have to ensure the interference is taken into account when scheduling different flows and allocating powers. The MCUH algorithm solves the joint problem scheduling and power allocation, for flow $k>1$, as follows:

\subsection*{Ordering:} Pick a subset of nodes (as potential relays)  and assign an ordering to those nodes. Let us call this ordered subset $\mathcal{I}_k =(1_k,2_k,...,j_k,...,n_k)$,  where $1_k$ corresponds to $s_k$ and $n_k$ corresponds to $d_k$. This set could for instance be obtained by picking the nodes that would have been picked if we were to run the single-flow algorithm of  section \ref{sec:multi_uni} for flow $k$. In \cite{Baghaie11, Baghaie12}, it was shown that Dijkstra's shortest path algorithm provides a good heuristic for ordering in a network with uniform distribution of nodes.

Notice that ordering does {\it not} dictate the time slot at which a node should transmit. It only states that any node that has decoded the message will only be {\it allowed} to retransmit when all nodes with smaller index have also decoded the message. Being {\it allowed} to transmit, also does not indicate that a node will in fact transmit.

Given this ordering, we now need to solve the joint scheduling and power allocation problem. 

\subsection*{Power Allocation:}

Let us call the power allocation algorithm, $PAM$ for power allocation multi-flow.  $PAM$, to be specified shortly, calculates the instantaneous optimal power allocation for flow $k$ at time $t$, given the set of instantaneous senders and receivers for flow $k$ and the set $\mathcal{B}(t)$ (of senders and receivers of flows $1$ to $k-1$ and their corresponding powers at that time slot). The design of $PAM$ is as follows:



$PAM$ is defined for flow $k$, to take as an input a set of transmitters ($\Psi_k(t)=\{1 ... i_k\}$) and a set of receivers ($\mathcal{R}_k(t)= \{i_k+1... j_k\}$), and the set of already scheduled nodes for that time-slot and their corresponding powers $\mathcal{B}(t)$, the channel between the nodes and the receiving threshold $\theta_k$. The objective of $PAM$ is to minimize the total sum power used to transmit flow $k$, by all the nodes transmitting that flow at time $t$. In other words, the objective is to minimize $ \sum_{q \in \Psi_k(t)} p^k_{qt}$, while satisfying a number of conditions. The output of this algorithm is the set of $p^k_{qt}$ for $q \in \Psi_k(t)$ and also the sum of powers, $\Omega^*_t(k)= \sum_{q \in \Psi_k(t)} p^k_{qt}$, which is the objective of the optimization. We also define a corresponding function of the same name $PAM$, that returns $\Omega^*_k(t) $ as its output for the corresponding input. In other words $PAM ( \Psi_k(t),\mathcal{R}_k(t),\theta_k, H, \mathcal{B}(t)) = \Omega^*_k(t).$ $PAM$ is described in Algorithm \ref{alg:pam}.

\begin{algorithm}[h!] 
  \caption{Power Allocation Multiflow (PAM) $(k,t)$}
  \label{alg:pam}
  \begin{algorithmic}[1]
   \STATE \textbf{INPUT:} $\Psi_k(t)$ and  $\mathcal{R}_k(t)$ for a given flow $k$ at time $t$,  $\mathcal{B}(t)$, $H$, $\theta_k$
     \STATE \textbf{OUTPUT:}   $p^k_{qt}$ for $q \in \Psi_k(t)$ and the objective value  $\Omega^*_k(t)$
    \STATE \textbf{Begin:}

    
    \IF{$q\in \mathcal{B}(t)$,  $\forall  q\in \mathcal{R}_k(t)$}
    
    \STATE  $\Omega^*_k(t) := \infty$.
   \STATE  \textbf{return} infeasible.
     
     \ELSE 
     \STATE {\begin{multline} \label{eqn:PAM}  \hspace{-1cm} \Omega^*_k(t) =\min \sum\limits_{q \in \Psi_k(t)} p^k_{qt} \hspace{0.8cm}  \textrm{s.t.} \\
\hspace{-5.8cm}1. \hspace{0.2cm}p^k_{qt} \geq 0,~~\forall q \in \Psi_k(t)   \\
\hspace{-1.8cm} 2.{ \tiny{\sum\limits_{q\in \Psi_k(t)} h_{qj} p^K_{qt} - \theta_k \left( N+\sum\limits_{u\in \Psi_f(t)} h_{uj}p^f_{ut}\right) \geq 0}}, \\ \hspace{5cm}   \substack{ \forall j \in \mathcal{R}_k(t)\\ ~~~~\forall f \in \{1,...,k-1\} } \\  
\hspace{-1.8cm} 3. {\tiny{\ \sum\limits_{v\in \Psi_f(t)} h_{vz} p^f_{vt} - \theta_f \left( N+ \sum\limits_{\substack{
   u\in \Psi_g(t) \\ g \neq f}}  h_{uz}p^g_{ut}  \right) \geq 0}},\\ \hspace{5cm}\substack{\forall z \in \mathcal{R}_f(t)\\ \forall g\in\{1,...,k\}\\ ~~~~\forall f \in\{1,...,k-1\}}\\
\hspace{-1cm}4. \hspace{0.2cm} p^k_{qt} = 0,~~~~  \forall q\in \mathcal{B}(t).\hspace{9cm}  \end{multline}   }
  
\STATE     \textbf{return} $\Omega^*_k(t)$,   $\{p^k_{qt}\}_{q \in \Psi_k(t)}$.
   \ENDIF

    \STATE \textbf{End}

  \end{algorithmic}
  \end{algorithm}

The $PAM$ algorithm works by first ensuring that a node cannot receive a message for flow $k$ at time $t$, if it has already been scheduled to participate in another flow in that time slot. This renders the power allocation task infeasible with the given set of transmitting and receiving nodes. In other words, it returns infinity and states the result to be infeasible. If that is not the case, $PAM$ proceeds to calculate $\Omega^*_k(t)$ by solving a linear optimization problem. In the optimization formulation, constraint $1$ ensures that there are no negative powers. Constraint $2$ ensures that the nodes assigned to receive flow $k$ at time $t$ will in fact accumulate enough energy to decode the message, despite the existing interference. Constraint $3$ ensures that the power being assigned to nodes in flow $k$, is not disturbing the previously scheduled flows.  Constraint $4$ ensures that a node cannot transmit a message for flow $k$ at time $t$, if it has already been scheduled to participate in another flow in that time slot.

\subsection*{Joint Scheduling and Power Allocation:}

Given this power allocation algorithm for flow $k$, and the ordering, all that remains to be done is scheduling. In other words, determining the set of transmitting and receiving nodes at each time slot  that need to be passed on to the power allocation algorithm. 

Given the ordered subset of $\mathcal{I}_k =(1_k,2_k,...,j_k,...,n_k)$,  we define $C_k(j_k,t)$ to be the minimum energy needed for flow $k$ to deliver the message to all the nodes up to node $j_k$ in $t$ steps or less.  We then use the following deterministic dynamic program to solve the joint scheduling and power allocation problems, optimally:

\begin{multline}\label{eq.twodimpolyMF}
C_k(j_k,t) = \min_{i_k \in (1,..,j_k)} [C_k(i_k,t-1) +...\\+ PAM(\{1 ... i_k\}, \{i_k+1... j_k\},\theta_k, H, \mathcal{B}(t))] 
\end{multline}
where $C_k(i_k,1) = PAM(1_k,\{2_k ... i_k\},\theta_k, H, \mathcal{B}(t))] $ \newline $\forall i_k \in \mathcal{I}_k \backslash 1_k$, and $C_k(1_k,t) = 0~~~\forall t$.  Thus, for flow $k$, the total minimum cost for covering $n_k$ nodes by time $T_k$ can be found by calculating $C_k(n_k, T_k)$.  

After each flow is scheduled, we save the set of scheduled transmitting nodes for each time-slot $t$ and their corresponding powers in $ \mathcal{B}(t)$. The algorithm is repeated for each flow $k>1$.

\section{Performance Evaluation}

In this section we compare the performance of the proposed heuristic against the analytical bounds for an example network with arbitrarily chosen three flows. We  also look at the effect of channel degradation in the overall performance.

We consider a network of  $100$ nodes uniformly distributed on a $20$ by $20$ square surface. The channels between all nodes are static, with independent and exponentially distributed channel gains (corresponding to Rayleigh fading), where $h_{ij}$ denotes the channel gain between node $i$ and $j$. The mean value of the channel between two nodes, $\overline{h_{ij}}$, is chosen to decay with the distance between the nodes, so that $\overline{h_{ij}} = d_{ij}^{-\eta}$, with $d_{ij}$ being the distance between nodes $i$ and $j$ and $\eta$ being the path loss exponent. The corresponding distribution for the channel gains is then given by
\[f_{h_{ij}}(h_{ij})= {\frac{1}{\overline{h_{ij}}}}\exp \left(\frac{h_{ij}(k)}{\overline{h_{ij}}}\right)\]

Notice that the minimum power calculated by different algorithms, shown on the y-axes of the graphs in this section, are normalized by value of $\theta$ (rendering it unit-less). 

\begin{figure}[h]
\scalebox{0.35} {

\includegraphics{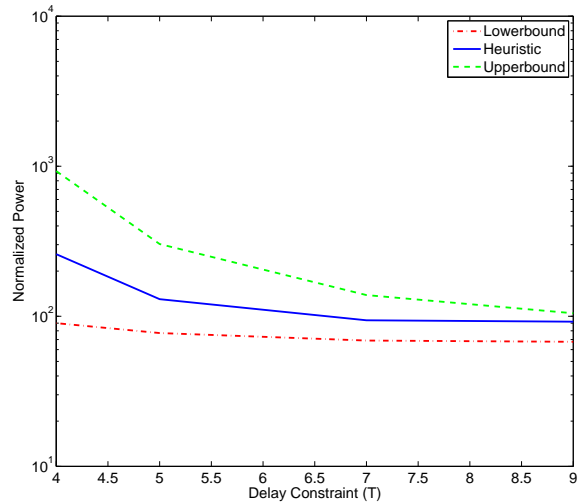} 
}
\caption{Performance of the heuristic against the analytical upper and lower bound.} 
\label{fig:100nodesbounds}
\end{figure}

Figure \ref{fig:100nodesbounds} shows the performance of the heuristic against that of the analytical bounds. As can be seen the heuristic is performing close to the lower bound.  Notice that the lower bound is an unachievable lower bound, in that  it assumes no interference is present. This means that its performance is not achievable by any algorithm. This is more emphasized when we have fewer time slots available, and thus we need to use more power to transmit the message creating a lot of interference that is ignored by the lower bound. As we get more time-slots available to us, the performance of the heuristic and the bounds seem to converge, which is what we expect as the solution goes to a multiplexing solution in all cases.

\begin{figure}[!t]
\scalebox{0.35} 
{
\includegraphics{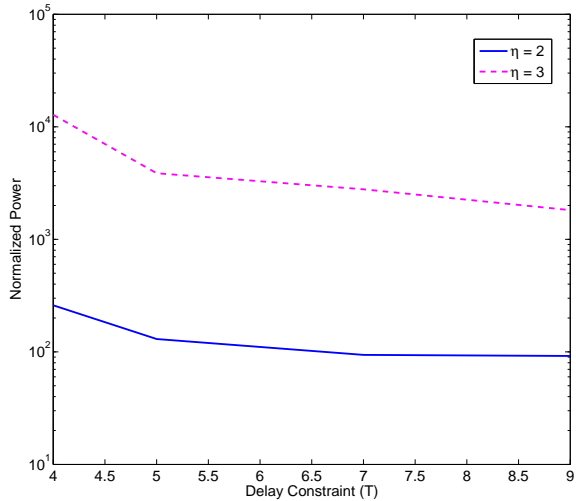} 
}
\caption{Effect of channel degradation on the total energy consumed.} 
\label{fig:Eta2_Eta3}
\end{figure}
We see the effect of poor channel conditions in Figure \ref{fig:Eta2_Eta3}. As expected the performance is degraded as the channel conditions become poor, this highlights the importance of having smart algorithms to minimize the energy consumption in such scenarios.

\section{Conclusion} \label{sec:multi_conc}

In this paper we formulated the problem of minimum energy cooperative transmission in a delay constrained multiflow multihop wireless network, as a combinatorial optimization problem, for a general setting of $k$-flows and formally proved that the problem is not only NP-hard but it is $o(n^{1/7-\epsilon})$ inapproxmiable. We proved inapproximability by reduction from the classic minimum graph coloring problem. To our knowledge, the results in this work provide the first such inapproxmiablity proof in the context of multiflow cooperative wireless networks.

We further proved that for the special case of $k=1$, the solution is a simple path and devised an optimal polynomial time algorithm for joint routing, scheduling and power control. We then used this algorithm to establish analytical upper and lower bounds for the optimal performance for the general case of $k$ flows, where the delay constraint is at least equal to $k$. Furthermore, we proposed a polynomial time heuristic for calculating the solution for the general case and evaluated the performance of this heuristic under different channel conditions and against the analytical upper and lower bounds.